\documentclass[12pt]{book}

\usepackage[dvips]{graphicx,color}
\usepackage{makeidx,tsukuba}

\makeauthorindex
\makeindex

\begin{document}

\BookTitle{\itshape The 28th International Cosmic Ray Conference}
\CopyRight{\copyright 2003 by Universal Academy Press, Inc.}
\pagenumbering{arabic}

\chapter{
Cosmic ray acceleration at parallel relativistic shocks in the presence 
of finite-amplitude magnetic field perturbations}

\author{%
%
%
Jacek Niemiec,$^1$ and Micha\l{} Ostrowski$^2$ \\
{\it (1) Institute of Nuclear Physics, ul. Radzikowskiego 152, 31-342 Krak\'ow,
 Poland\\
(2) Astronomical Observatory, Jagiellonian University, ul. Orla 171, 30-244 
Krak\'ow, Poland} \\
}

\section*{Abstract}
We study the first-order Fermi acceleration process at parallel shock
waves by means of Monte Carlo simulations. A 'realistic' model of the
magnetic field turbulence is applied involving sinusoidal waves imposed
on the background mean field parallel to the shock normal. The
finite-amplitude magnetic field perturbations lead to locally oblique
field configurations at the shock and the respective magnetic field
compression. It results in modification of the particle acceleration
process introducing some features observed in oblique shocks, e.g.
particle reflections from the shock. For the parallel mildly relativistic shocks
we demonstrate for the first time a
(non-monotonic) variation of the accelerated particle spectral index
with the turbulence amplitude.

\section{Introduction}
As discussed by Ostrowski [9] for nonrelativistic shocks, the presence of
finite-amplitude magnetic field perturbations modify character of the diffusive 
particle acceleration at the shock wave with the mean 
field parallel to the shock normal. The effect arises due to locally oblique 
field configurations formed by long-wave perturbations at the shock front and 
the respective magnetic field compressions. As a result the mean particle energy
gains may increase and the particles reflected from the shock front may occur. 
The same phenomena should occur at relativistic shocks [cf. 10].

In the simplified
numerical simulations of the first-order Fermi acceleration at parallel
mildly relativistic shocks the
acceleration time scale reduces with increasing turbulence level, but no
spectral index variation occurs [1, see 2 for ultrarelativistic shocks].
However, the considered acceleration models apply very simple
modeling of the perturbed magnetic field effects by introducing particle
pitch-angle scattering.
The purpose of the present work is to simulate the first-order Fermi
acceleration process at mildly relativistic shock waves propagating  in  
more realistic perturbed magnetic fields, including a wide wave vector
range turbulence with the power-law spectrum. The magnetic 
field is continuous across the shock, according to the respective jump
conditions. This feature leads to substantial modifications of the acceleration
process at parallel shocks.

Below the upstream (downstream) quantities are labeled with 
the index `1' (`2').

\section{Simulations}
In the simulations trajectories of ultrarelativistic test particles are derived
by integrating their equations of motion in the perturbed magnetic field
[for details see: Niemiec, Ostrowski, in preparation].
A relativistic shock wave is modeled as a plane discontinuity 
propagating in electron-proton plasma. The magnetic field is defined upstream
of the shock. It consists of the uniform component, $\vec{B}_{0,1}$, 
parallel to the shock normal and finite-amplitude perturbations imposed upon it.
The perturbations are modeled as a superposition of 294 sinusoidal static waves 
of finite amplitudes [cf. 10]. They
have either a flat $(F(k)\sim k^{-1})$ or a Kolmogorov 
$(F(k)\sim k^{-5/3})$ wave power spectrum in the (wide) wave vector range 
$(k_{min}, k_{max})$, where $k_{max}/k_{min}=10^5$.
The shock moves with the velocity 
$u_1$ with respect to the upstream plasma. The downstream flow 
velocity $u_2$ and the magnetic field structure are obtained from the
hydrodynamic shock jump conditions. Derivation of the shock compression ratio  
as measured  in the shock rest frame, $R = u_1/u_2$, is
based on the approximate formulae derived in Ref. [3].
In the analysis of the acceleration process the particle  radiative (or other) 
losses are neglected.

\section{Results for parallel shocks}
In Fig. 1 we present particle spectra for the parallel shock wave
with  $u_1 = 0.5c$. The shock compression ratio is $R = 5.11$. 
The particle spectra are measured at the shock for three different magnetic
field perturbation amplitudes and the flat (Fig. 1a) or the Kolmogorov 
(Fig. 1b) wave power spectrum.
One can note that the particle spectral indices deviate from the small 
amplitude results of the pitch angle scattering model [3-6]. In addition, the  
increasing magnetic field perturbations can produce non-monotonic changes of the
particle spectral index -- the feature which has not been discussed for parallel 
shocks so far. 
Analogously to oblique shock waves [cf.  an accompanying paper in this
volume], our particle spectra  are non power-law in the full
energy range and the shape of the spectrum vary with the amplitude of
turbulence and the wave power index. 

The non-monotonic variation of the spectral index with the turbulence amplitude
results from modifications of the particle acceleration process at the shock.
The long-wave finite-amplitude perturbations produce locally oblique
mag-\newpage
\begin{figure}[th]
\begin{center}
\includegraphics[scale=0.79]{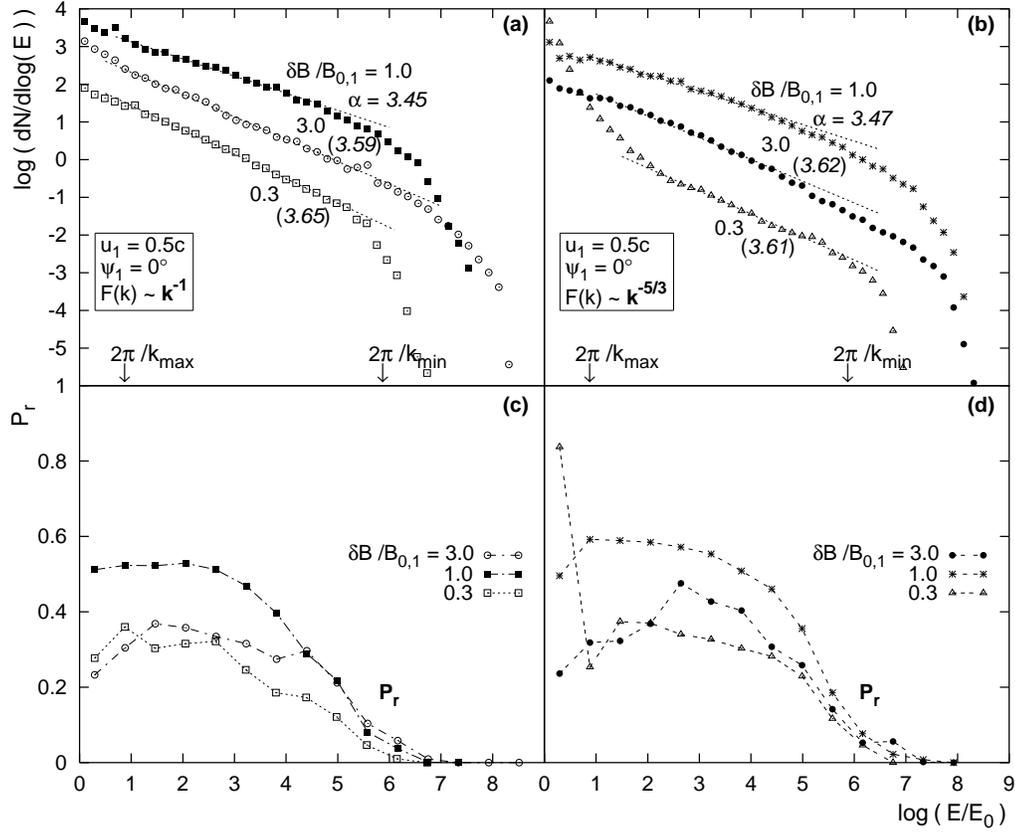} 
\end{center}
\caption{ Accelerated particle spectra at the parallel shock wave
in the shock rest frame for (a)
the flat ($F(k) \propto k^{-1}$) and (b) the Kolmogorov ($F(k) \propto k^{-5/3}$)
wave spectrum of magnetic field perturbations. 
The upstream perturbation amplitude $\delta B / B_{0,1}$
is given near the respective results. Linear fits to the power-law parts of the 
spectra are presented and values of the phase space distribution function 
spectral indices $\alpha$ are given in parentheses. Particles in the energy 
range indicated by arrows can effectively interact with the magnetic field
inhomogeneities ($k_{min} < k_{res} < k_{max}$). For upstream particles
probabilities of reflection from the shock, $P_r$, are presented as a function 
of particle energy for the respective particle spectra above (the transmission 
probability $P_{12}=1-P_r$).} 
\end{figure}
\noindent netic field
configurations and lead to occurrence of particles reflected from the
compressed field downstream of the shock. 
Probability of reflection depends on the turbulence amplitude and the amount of
field perturbations with wavelengths larger than the resonance wavelength for a
given particle, as presented in
Figs. 1c and 1d. For $\delta B / B_{0,1}=1.0$ the reflection probability is
higher as compared to the other  perturbation amplitudes considered and the
particle spectrum is flatter. For the chosen by us smaller 
($\delta B/B_{0,1}=0.3$) and larger 
($\delta B / B_{0,1}=3.0$) turbulence amplitudes the reflection and
transmission probability do not differ considerably, which results in the
similar values of the spectral indices. One can note in this place, that the
obtained spectra for the Kolmogorov case seem to exhibit a continuous slow 
change of inclinations. 
Thus the fitted power-laws depend to some extent on the energy
range chosen for the fit. One can also note a steep part of the spectrum at low
energies for $\delta B / B_{0,1}=0.3$ in Fig. 1b.

The presented reflection (transmission) probabilities decrease (increase) 
at high particle energies due to a limited dynamic range of the magnetic 
field turbulence. 
The locally oblique field configurations are mainly formed by long-wave 
perturbations ($k < k_{res}$) [cf. 9]. For high energy particles with
$k_{res} < k_{min}$ there are no respectively long waves and the upstream 
particles can be only transmitted downstream of the shock. In these conditions
the acceleration process would converge to the `classic' parallel shock 
acceleration model, but in our simulations particles move far to the introduced
escape boundary forming a cut-off.

\section{Summary}
The simulations of the first-order 
Fermi acceleration process acting at  parallel relativistic shock waves are
presented. In the presence of finite-amplitude perturbations the particle 
spectral indices vary in a non-monotonic way 
with the turbulence amplitude due to non trivial character of the particle
interaction with the shock, including particle reflections. This feature is also expected to lead to some 
decrease of the particle acceleration timescale in comparison to previous 
estimates [7, cf. 8].

 The work was supported by the Polish State 
Committee for Scientific Research in 2002-2004 as a research project 2 P03D 008
23 (JN) and in 2002-2005 as a solicitated research project 
PBZ-KBN-054/P03/2001 (M0).

\section*{References}
\noindent
1. Bednarz J., Ostrowski M. 1996, MNRAS 283, 447\\
2. Bednarz J., Ostrowski M. 1998, Phys. Rev.Lett. 80, 3911\\ 
3. Heavens A., Drury L'O.C. 1988, MNRAS 235, 997\\
4. Kirk J.G., Heavens A. 1989,  MNRAS 239, 995\\
5. Kirk J.G., Schneider P. 1987, ApJ 315, 425\\
6. Kirk J.G., Schneider P. 1987, ApJ 322, 256\\
7. Lagage P.0., Cesarsky C.J. 1983, A\&Ap 118, 223\\
8. Ostrowski M. 1988, MNRAS 233, 257\\
9. Ostrowski M. 1988, A\&Ap 206, 169\\
10. Ostrowski M. 1993, MNRAS 264, 248\\
\endofpaper
\end{document}